\documentclass[12pt]{article}

\usepackage{cite}
\usepackage{times}
\usepackage{amsmath}
\usepackage{graphicx}

\newenvironment{sciabstract}{%
\begin{quote} \bf}
{\end{quote}}

\title{Continuous-wave nitrogen-vacancy diamond laser system assisted by a red diode laser}

\author
{Lukas Lindner,$^{1\ast}$ Felix A. Hahl,$^{1}$ Tingpeng Luo,$^{1}$ Guillermo Nava Antonio,$^{1}$\\ Xavier Vidal,$^{1}$ Marcel Rattunde,$^{1}$ Takeshi Ohshima,$^{2}$ Marco Capelli,$^{3}$\\ Brant C. Gibson,$^{4}$ Andrew D. Greentree,$^{4}$ Rüdiger Quay,$^{1}$  and Jan Jeske$^{1\ast}$\\
\\
\normalsize{$^{1}$Fraunhofer IAF, Fraunhofer Institute for Applied Solid State Physics}\\
\normalsize{$^{2}$Department of Materials Science, Tohoku University}\\
\normalsize{$^{3}$School of Science, RMIT University}\\
\normalsize{$^{4}$ARC Centre of Excellence for Nanoscale BioPhotonics,}\\
\normalsize{ School of Science, RMIT University}\\
\\
\normalsize{$^\ast$To whom correspondence should be addressed;}\\
\normalsize{E-mail: lukas.lindner@iaf.fraunhofer}\\ \normalsize{E-mail: jan.jeske@iaf.fraunhofer.de.}
}

\date{}

\begin{document} 




\maketitle


\begin{sciabstract}
 Diamond has long been identified as a potential host material for laser applications. This potential arises due to its exceptional thermal properties, ultra-wide bandgap, and color centers which promise gain across the visible spectrum. More recently, coherent laser methods offer new approaches to magnetometry. However, diamond fabrication is difficult in comparison to other crystalline matrices, and many optical loss channels are not yet understood. Here, we demonstrate the first continuous-wave nitrogen-vacancy (NV) color center laser system. To achieve this, we constructed a laser cavity with both, an NV-diamond medium and an intra-cavity anti-reflection coated diode laser. This dual-medium approach compensates intrinsic losses of the cavity by providing a fixed additional gain below threshold of the diode laser. We observe the first clear continuous-wave laser threshold in the output of the laser system as well as linewidth narrowing with increasing green pump power on the NV centers. Our results are a major development towards coherent approaches to magnetometry.
\end{sciabstract}


\section{Introduction}
Diamond is an interesting material for lasers because of its wide band gap, high optical transparency and thermal conductivity, both as an active medium for Raman lasing \cite{williams2018high} and a well-suited host for color centers as a laser medium\cite{Rand.1985, rand1994damiandlasers, Dobrinets.2013}. One of the most famous and well-characterized color centers is the negatively charged nitrogen vacancy (NV) center in diamond, due to its atom-like energy level structure and long coherence time at room temperature. NV centers are used in a wide range of applications from quantum information\cite{wrachtrup2006processing} to sensing applications for temperature \cite{neumann2013high}, orientation and angular velocity \cite{soshenko2021nuclear}, magnetic signature of biological samples\cite{hall2012high}, and electric\cite{dolde2011electric} or magnetic fields \cite{maze2008nanoscale, ruan2018magnetically, simpson2016magneto, Hall.2012}.


Most of these sensor concepts measure the response of the fluorescence signal to environmental parameters. The magnetic sensing approach is based on the Zeeman-effect of the energy levels of the NV centers. 
The precision of fluorescence readout depends on the amount of collected light and its response to the measured quantity, i.e. the measurement contrast.
There are different approaches to increase the collection efficiency such as microlenses \cite{robledo2011high} or microcavities and waveguides for better coupling \cite{eisenach2021cavity, hausmann2012integrated}. 
A significant improvement of the sensitivity of magnetic field measurements can be achieved by laser threshold magnetometry  (LTM) \cite{jeske2016laser}. This measurement concept benefits from an increased intensity through stimulated emission and an improved collection efficiency due to its coherent output. Due to the nonlinearity of the laser threshold, the measurement contrast is also enhanced. 
Theoretically, LTM has been projected to reach sensitivities as low as \textasciitilde 1\,fT/$\sqrt{\text{Hz}}$ \cite{jeske2016laser}. In comparison, the highest demonstrated sensitivity with a fluorescence based approach has now reached the regime of sub pT/$\sqrt{\text{Hz}}$ \cite{zhang2020diamond, fescenko2020diamond, Barry.10.05.2023, Graham.2023}. This indicates the great potential of the concept of LTM for NV magnetometry, but also for other applications.

Different realizations of LTM systems have been proposed. 
Dumeige \textit{et al.}~proposed an implementation using an NV-diamond as an optical absorber within an external laser cavity with an additional active laser medium. They project a theoretical sensitivity of 700\,fT/$\sqrt{\text{Hz}}$ \cite{dumeige2019infrared}.
Another theoretical approach uses the absorption of the pump laser within an external Fabry-Perot cavity. Hereby, the sensitivity can be estimated to 300-20\,fT/$\sqrt{\text{Hz}}$ \cite{webb2021laser, ahmadi2018nitrogen}. 
Nair \textit{et al.}~report on a theoretical investigation of an NV-diamond Raman laser and calculate a sensitivity of 1.62\,pT/$\sqrt{\text{Hz}}$ \cite{nair2021absorptive}.

Experimentally, stimulated emission from NV centers was demonstrated in 2017 \cite{jeske2017stimulated}, followed by a demonstration of NV-light amplification in a fiber cavity \cite{nair2020amplification}. Material characterization and optimization \cite{fraczek2017laser, Luo.2022,Luo.2023} were necessary to enable further improvements in material and cavity design. Hahl \textit{et al.} demonstrated light amplification in an externally seeded cavity using a red seeding laser. Magnetic field-dependence with a record contrast and a projected sensitivity of 15\,pT/$\sqrt{\text{Hz}}$ were measured, demonstrating a one-order-of-magnitude improvement in sensitivity over spontaneous emission fluorescence sensing \cite{Hahl.2022}.

In 2021, Savvin \textit{et al.}~demonstrated a pulsed NV laser system, wherein the NV centers provide the laser system gain \cite{savvin2021nv, Lipatov.2022, Mironov.2023}.
For sensing applications, the laser pulses have the significant disadvantage of a lack of signal between laser pulses. A continuous-wave (cw) laser can increase the signal intensity, sensitivity and stability of a measurement significantly. For equal laser gain, the signal strength of a cw laser is higher than that of a pulsed laser, by a factor corresponding to the ratio of pulse duration to repetition time. In addition, a cw read-out is better suited for most sensing measurements concepts. A cw laser is thus essential for sensing with improved sensitivity.

Until now, in all realizations it was not possible to show a laser system wherein the NV centers are pumped with a continuous-wave optical laser and therefore the stimulated emission of the NV centers themselves may be used as a sensor. This is due to several difficulties concerning NV-doped diamond as a laser-active medium: First, while diamond is a sufficiently transparent medium for optical wavelengths, NV-doped samples often show significant absorption around 700\,nm \cite{fraczek2017laser} at the maximum of the fluorescence~\cite{luo2022creation}.
Second, the NV center has two possible charge states, of which only the NV$^-$ is suitable for sensing applications. The proportion of NV$^-$ and NV$^0$ depends on the excitation power. Especially at high powers, photoionization drives a transition from NV$^-$ to NV$^0$. Thirdly, high powers (at pump or read-out wavelength) within the cavity lead to an "induced absorption" effect \cite{Hahl.2022, savvin2021nv, nair2020amplification}. 
The absorption in NV-diamonds could be overcome by an effective increase of the laser threshold of NV-doped diamond as an active medium, however the photoionization and induced absorption effects limit the maximal pump power on the NV-diamond which can be used to overcome its laser threshold.

Here, we propose a new solution: Decreasing the laser threshold of the NV laser system by creating optical gain at the same wavelength range within the same cavity. 
The concept relies on a second active laser medium to compensate the intrinsic losses of the cavity and the diamond, so that the NV centers only have to add a small gain to overcome the laser threshold. We present an experimental realization using a linear cavity containing a laser diode and an NV-doped diamond sample. With this, we demonstrate a cw-laser-threshold of an NV laser system for the first time, which we believe is an essential step toward an LTM sensor system.

\section{Diode-assisted NV-laser system}
\subsection{Energy level scheme}
\begin{figure}
	\includegraphics[width=\textwidth]{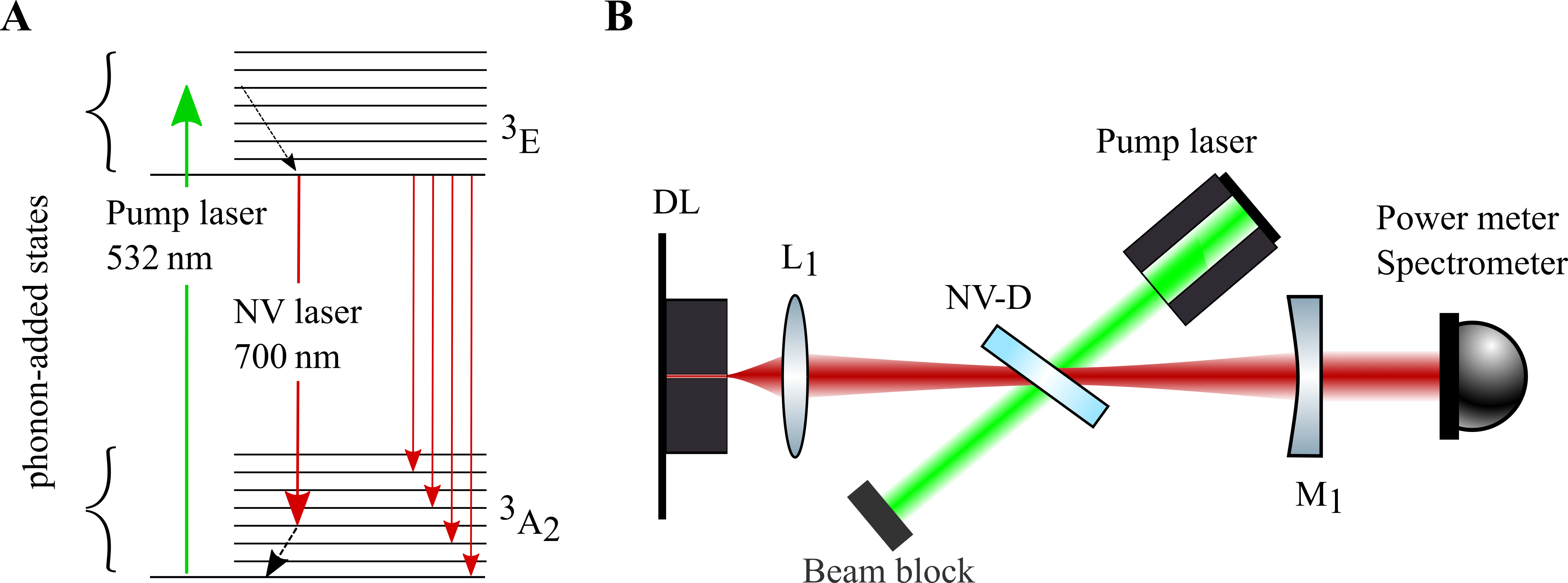}
	\caption{\textbf{Experimental design.} \textbf{(A)} Energy levels of the NV center:  the green pump laser addresses a transition from the ground state to a phonon-added excited state, the laser transition is from the excited state to a phonon-added ground state.
	\textbf{(B)} Sketch of the linear cavity setup:  The backside mirror of the diode laser (DL) and the external outcoupling mirror (M$_1$) form the cavity, the diode laser and the NV diamond (NV-D) provide gain, the lens (L$_1$) is necessary to focus the output of the anti-reflection coated side of the diode laser. \label{fig:setup}}
\end{figure}

An NV laser system was proposed by Rand \textit{et al.} in 1994\cite{rand1994damiandlasers}. The energy levels relevant to this work are shown in figure \ref{fig:setup}a). The lowest energy level is that of the electron ground state ($^3$A$_2$). The first excited level is within the phonon-added states of the excited electron state $^3$E. The transition can be excited with a 532\,nm laser (pump laser). The energy states in the phonon-added excited states relax almost instantaneously to the lowest excited state without phonons, which is the upper laser level. The lower laser level is a phonon-added electronic ground state, which then in turn relaxes rapidly to the lowest phonon-added state. The two electronic states and their phonon-added states form a 4-level-laser system, which allows for lasing without the need to pump the majority of the population to the excited state to achieve population inversion, since the phonon-added ground state is de-populated quickly. This 4-level laser also avoids strong photoionization toward NV$^0$ which would occur for strong light fields around the zero-phonon line of 637\,nm.

\subsection{Experimental realization}
To realize a cw NV laser system, we combine the NV centers as a gain medium with an external-cavity diode laser as an additional gain medium. The linear cavity setup is shown in figure \ref{fig:setup}b). The diode laser (DL, Sacher Lasertechnik SAL-0690-025) is anti-reflection coated and emits around 690\,nm wavelength.
This wavelength range was chosen as a compromise between highest optical gain from the NV centers with low absorption losses of the NV-diamond sample \cite{fraczek2017laser} and technical availability of diode lasers.

The lens L$_1$ (2.75\,mm focal length) collects and refocuses the diode emission. At the second focal point, the NV-diamond sample is placed. The sample is a high-pressure-high-temperature (HPHT) sample, which was irradiated with an electron beam and annealed \textit{in situ}, according to the process described by  Capelli \textit{et al.} \cite{capelli2019increased}. The final NV concentration is about 2\,ppm. The sample has a thickness of around 350\,\textmu m. The NV-diamond is placed in its Brewster's angle to minimize reflection losses. The emission from the laser diode shows a longitudinal multimode structure from which we can select one by introducing a small misalignment of the diamond sample to its Brewster's angle, creating a residual etalon effect.

The outcoupling mirror M$_1$ has a reflectivity of around 95\,\% and a radius of curvature of r=-50\,mm.
For laser beam diagnostics, we use a power meter (Thorlabs S130C) with an accuracy of 3\,\% and a fiber-coupled spectrometer (Princton Instrument, HRS-750) with a spectral resolution of 0.02\,nm.

In the following experiments, both active media of the laser system will be operated near their laser threshold. To this end, the NV-diamond sample is pumped with a green laser with 532\,nm emission wavelength.

\section{Experimental results}
\begin{figure}[hpt]
	\centering\includegraphics{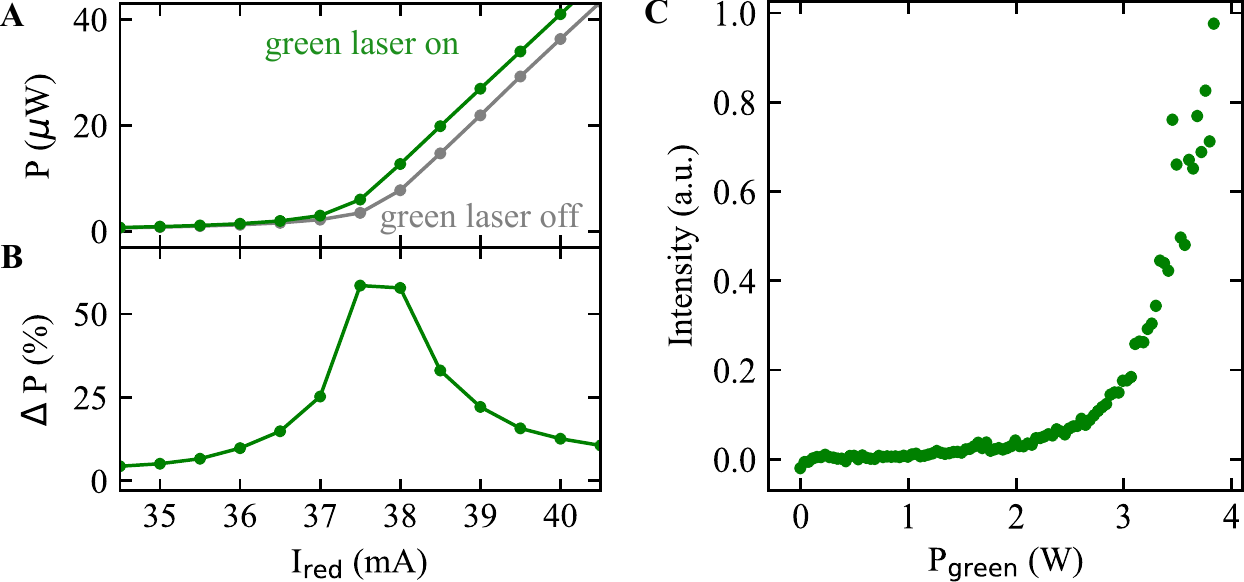}
	\caption{\textbf{Output power of the hybrid laser system.}
	\textbf{(A)} Output power (P) of the laser system as a function of the laser diode current (I$_\text{red}$). Gray: without NV gain (green pump laser turned off), green: with fixed NV gain.
	\textbf{(B)} Relative change in output power ($\Delta$P) between the two curves shown in a).
	\textbf{(C)} Demonstration of a cw NV laser threshold: Output of the NV laser system as a function of green pump power on the NV-diamond sample. The laser diode is operated with a fixed diode current below laser threshold, which is then overcome by the NV gain.   
	\label{fig:uiplot}}
\end{figure}

\subsection{Diode laser characteristics with fixed NV gain}
As a first step, the laser threshold of the diode laser will be determined, while the diamond is inserted into the cavity but not pumped (green pump laser turned off). The gray curve of Fig.~\ref{fig:uiplot}~a) shows the output power of the diode laser as a function of the laser diode current. 
The output power was measured with the power meter. The laser threshold is clearly visible at a diode current of 37.45\,mA.
In the next step, the measurement is repeated with the diamond pumped with 532\,nm green laser light (Fig.~\ref{fig:uiplot}a), green curve).
The output of the laser system is higher when the NV centers are pumped, which we assign to additional stimulated emission from the NV centers. The shift of the laser threshold shows an increase in round-trip gain, as the fixed cavity losses are overcome at lower diode pumping. Above threshold, the curves with and without the green pump laser show comparable slopes. This is in line with a fixed increase in gain due to constant pumping of the NV centers, which simply shifts the whole curve to the left. Figure \ref{fig:uiplot}b) shows the relative change ($\Delta$ P) in output power between the two curves, which was calculated as the ratio of the difference between the two curves to the measured power without green laser pumping. The influence of the optical state of the NV-diamond is largest at the laser threshold, where the green pump laser leads to an increase of output power by more than 50\,\%. This confirms the principle that cavity enhancement of NV readout is optimal at or just above the laser threshold, which was theoretically predicted in \cite{jeske2016laser}.  

To summarize, we have demonstrated that the threshold of the diode laser is shifted by an additional NV center gain. This result is the first cw measurement of stimulated emission of NV centers without an external seeding laser.

\begin{figure}[t!] 
	\centering\includegraphics{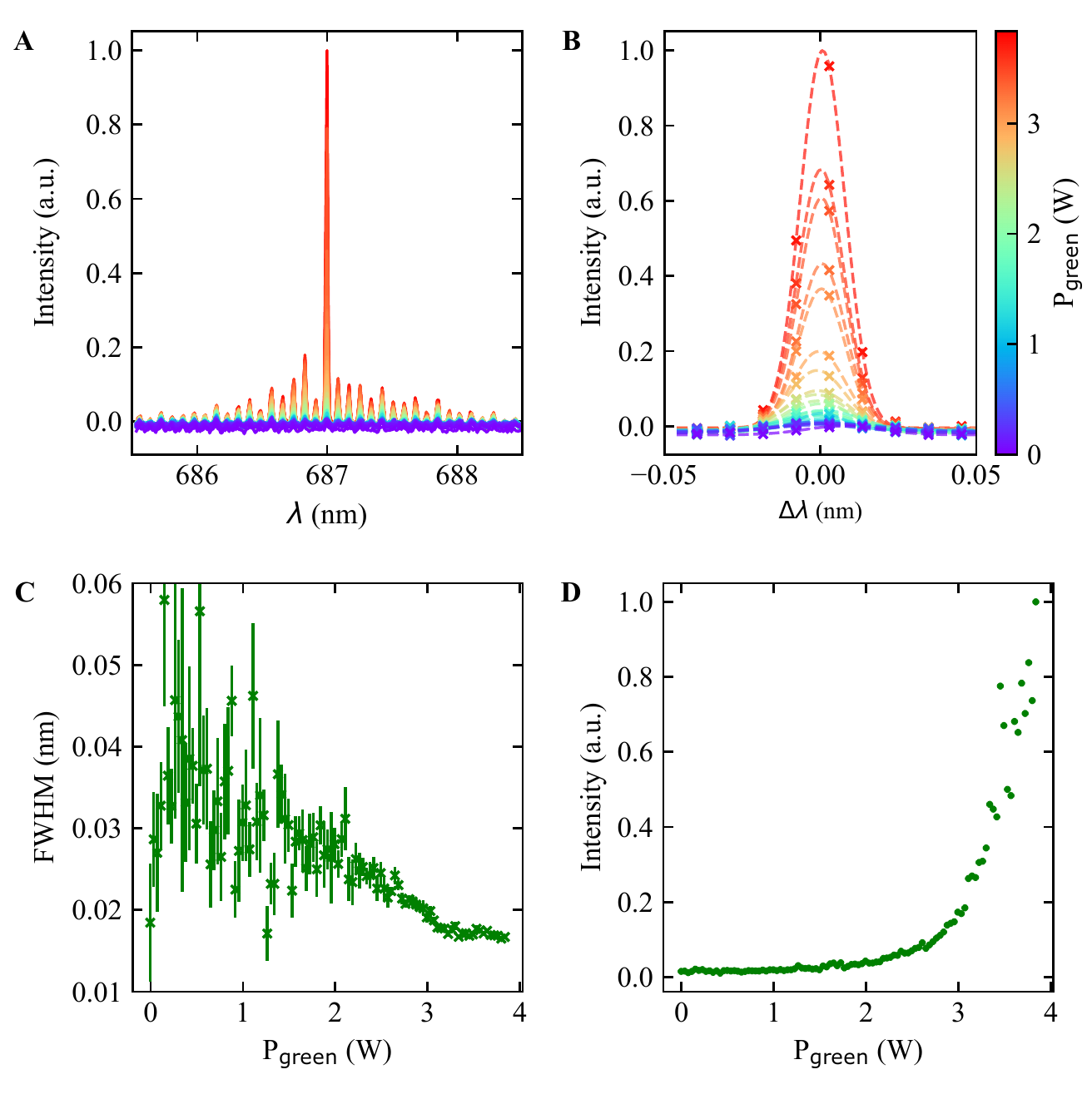}
	\caption{
		\textbf{Spectral output and linewidth narrowing of the NV laser system.}
		\textbf{(A)} Spectra of the diode-assisted NV laser system. The color corresponds to the green pump power (color scale on the right).
		\textbf{(B)} Detailed view of the main laser line (measured values shown as crosses) with a fitted Gaussian profile (dashed).
		\textbf{(C)} Laser linewidth (full width at half maximum, obtained from the fitted Gaussian curves) as a function of pump power. A reduction of the linewidth above the threshold (at $\sim$2.5\,W) is clearly visible.
		\textbf{(D)} Maximum amplitude from the Gaussian fit of the laser line as a function of pump power. The laser threshold at $\sim$2.5\,W matches with the spectrally averaged data in Fig~\ref{fig:uiplot}c).
	} \label{fig:laserspectrum}
\end{figure}

\subsection{NV laser characteristics with fixed diode gain}

Next, we aim to achieve an NV center laser which is only assisted by the diode, i.e. a laser threshold measured as a function of the NV centers' pump power, rather than the electric pumping of the diode. For this purpose, we set the diode current to a fixed value below threshold. This way it contributes gain which compensates for cavity round-trip losses, but the actual laser operation is then achieved by the NV-diamond. To ensure that we only measure coherent laser light (in the regime of a few \textmu W) and reject the fluorescence light, we apply spectral filtering by measuring with a spectrometer and integrating the counts within a narrow spectral window around the laser line at 687\,nm, with a bandwidth of 0.04\,nm. This is in addition to a spatial mode selection, as most of the fluorescence light of the NV centers can be assumed to be rejected by the small numerical aperture of the spectrometer (and the imaging optics towards it). Here one advantage of the laser readout over fluorescence becomes apparent: the signal is entirely delivered to the detector via the collimated laser beam, while fluorescence is emitted into all directions.

The resulting laser output as a function of the NV centers' pump power (P$_{\text{green}}$) is shown in Figure \ref{fig:uiplot} c).  
The laser output spectral power clearly shows a typical laser characteristic with a laser threshold. Below the threshold power of around 2.5\,W the output is nearly constant with a low slope, while above the threshold a linear increase is seen, with a much higher slope. The threshold appears washed out because of the high intra-cavity losses and possible instabilities in the laser diode current \cite{siegman1986lasers}.

The laser output measurement shows significant noise, particularly above threshold, which is common in laser systems. At threshold, the laser diode gain is in balance with all intra-cavity losses. Since the laser diode gain is likely much higher than the gain of the NV centers, a small variation of the system changes the output of the laser system rapidly, which results in fluctuations.
Overall the results indicate that the NV centers contribute to the NV-laser system in cw lasing operation.

\subsection{Spectral characteristic}

Further proof of laser action can be shown via the behavior of the emission linewidth around the threshold.
Light created by stimulated emission amplifies the laser mode with a spectral characteristic: As the stimulated emission rate is proportional to the light intensity of the mode, the intense spectral center of the mode is amplified more strongly than its edge.
This leads to an effective narrowing of the emission line when lasing occurs with increasing pump power \cite{Faizani.2008, hide1996semiconducting}. We investigate this behavior in our system.

For this, we analyze the measured spectra as a function of green pump power. Each individual spectrum is shown in Figure \ref{fig:laserspectrum} a) with a color corresponding to the  pump power of the green laser on the NV-diamond sample. 
Multiple resonance lines can be seen, with the line at 687\,nm being clearly dominant, especially at higher pump powers. The resonant frequencies correspond to the mode profile of the laser diode (residual etalon effect). Figure \ref{fig:laserspectrum} b) shows a zoom in around the main laser line at 687\,nm. We fitted each spectrum with a Gaussian curve (dashed line). The full width at half maximum (FWHM) obtained from the fit is shown in Figure \ref{fig:laserspectrum} c) and the amplitude is shown in d), both as a function of the green pump power on the NV-diamond sample. The fitted amplitude of the laser line is similar to the integrated spectral intensity as shown in Figure \ref{fig:uiplot} c). 
Below laser threshold, the spectrum shows strong fluctuations, which results in deviations of the laser linewidth.
At pump powers above 2\,W, the laser linewidth shows a steady decline and reaches a value of about 0.02\,nm, which is the resolution limit of the spectrometer. The reduction of the laser linewidth above threshold is caused by the NV centers, as the laser diode current is kept constant during the measurement. This narrowing of the linewidth clearly shows the transition from spontaneous to stimulated emission at the laser threshold and furthermore confirms the on-set of self-sustained CW lasing operation of the NV centers.

\section{Conclusion}
We have demonstrated a continuous-wave NV laser system using an external-cavity diode laser as an additional gain medium to provide fixed additional gain which off-sets cavity losses and is operated below threshold. The laser diode defines the wavelength where NV optical gain can create a self-sustained cw laser. We have shown that the gain of NV centers lowers the laser threshold of the red diode laser if the NV centers are pumped with a green laser. A high contrast of more than 50\,\% of the diode laser system with or without NV gain was demonstrated. 

We then further explored the cw NV laser system by varying the NV centers' pump power, while keeping the diode laser current at a fixed value below threshold. This NV laser demonstrated a  cw laser threshold in its spectral output power. We also observed the typical narrowing of the laser linewidth with increasing laser intensity.

Our results provide the step from pulsed to cw lasing and from externally seeded to self-sustained NV lasing with fixed-gain assistance by the diode laser. This shows that the NV center is the second diamond color center (after the H3 or NVN center \cite{DeShazer.1985} ) which can be used as a cw laser medium. This provides a basis for using the optical non-linearity of the laser cavity to enhance NV sensing and promises new sensitivity records as the self-seeding allows for high contrast and signal intensity.
In addition to LTM, other applications of NV-based quantum sensing, which are now typically based on fluorescence, may benefit from a coherent read-out of stimulated emission from NV centers.

\section*{Acknowledgements}
We thank the company Sacher Laser for the assistance with the diode.

\paragraph*{Funding:}
L.L. and J.J. acknowledge funding from the German federal ministry for education and research, Bundesministerium für Bildung und Forschung (BMBF) under grant no. 13XP5063 and 13N16485. B.C.G. and A.D.G. acknowledge funding from the U.S. Office of Naval Research Global (ONRG) Global-X Challenge (N62909-20-1-2077), Asian Office of Aerospace Research and Development (FA2386-18-1-4056), and the Australian Research Council (ARC) Centre of Excellence for Nanoscale BioPhotonics (CE140100003).

\paragraph*{Author contributions:}
L.L., F.A.H., X.V. and J.J. conceived the idea of the cavity setup, L.L. and G.N.A. prepared and performed the experimental setup,  L.L. analysed the data, M.R., F.A.H. and X.V. supported in building the experimental setup and with helpful discussions about the setup development. T.O. performed electron irradiation and annealing for NV-creation of the sample. T.L. helped characterising the samples, L.L., F.A.H., X.V, M.R., M.C., R.Q., J.J., A.D.G. and B.C.G. discussed and interpreted the results and provided feedback to the manuscript. L.L. wrote the manuscript with the support of J.J..

\paragraph*{Competing interests:}
F.A.H., J.J., T.L. hold a patent related to the content filed by Fraunhofer-Gesellschaft zur Förderung der
angewandten Forschung e.V. (102021209666.2, 02.09.2021).
Otherwise the authors declare that they have no competing interests.
\paragraph*{Data and materials availability:}
All data needed to evaluate the conclusions in the paper are present in the paper. Additional data related to this paper may be requested from the authors

\bibliography{scibib}
\bibliographystyle{ieeetr}

\end{document}